\documentclass[prl,twocolumn,showpacs,nobibnotes,floatfix,superscriptaddress]{revtex4}
\usepackage{latexsym,color,amssymb,amsmath,epsfig}

\newcommand{\opr}[1]{\operatorname{#1}}

\begin{document}

\title{A Universal Operator Theoretic Framework for Quantum Fault Tolerance}
\author{Gerald Gilbert, Michael Hamrick, Yaakov S. Weinstein \\
{\it {\small Quantum Information Science Group, {\sc Mitre}, 260
Industrial Way West,
Eatontown, NJ 07224, USA}} \\[.05in]
Vaneet Aggarwal, A. Robert Calderbank \\
{\it {\small Department of Electrical Engineering, Princeton
University, Princeton, NJ 08544, USA}}}

\begin{abstract}
In this paper we introduce a universal operator theoretic framework
for quantum fault tolerance.  This incorporates a top-down approach
that implements a system-level criterion based on specification of
the full system dynamics, applied at every level of error correction
concatenation. This leads to more accurate determinations of error
thresholds than could previously be obtained. This is demonstrated both formally and with an explicit numerical example. The basis for our approach is the Quantum Computer Condition (QCC), an inequality governing the evolution of a quantum computer. We show that all known coding schemes are actually special cases of the QCC. We demonstrate this by introducing a new, operator theoretic form of entanglement assisted quantum error correction, which incorporates as special cases all known error correcting protocols, and is itself a special case of the QCC.
\end{abstract}

\maketitle

\section{Introduction}\label{sec:intro}
The theory of quantum computational fault tolerance allows for
successful quantum computation despite faults in the implementation
of the quantum computing machine \cite{Shor96}. To achieve fault
tolerance, quantum information is encoded into logical qubits
comprised of a suitable number of physical qubits, and quantum gates
are implemented in such a way that a single error affects only one
of a set of encoded qubits. Quantum error correcting techniques
\cite{QEC} are then applied, making use of syndrome measurements, to
correct errors that may have occurred. To reduce the probability of
multiple errors occurring during any single gate operation, quantum
error correction methods (or error avoidance methods, such as
decoherence free subspaces \cite{DFS} or noiseless subsystems
\cite{NS}) are concatenated \cite{KL}, thus allowing for the
possibility of fault-tolerant computation.

In order to successfully implement fault-tolerant quantum computer
operation, it is essential to have accurate values of error
thresholds. Thus, the determination of these values has been an
important research focus over the past decade \cite{Error}. The
standard treatment of fault-tolerance thus far presented in the
literature \cite{std_qft} may be characterized as a ``bottom-up"
approach. In this standard, bottom-up approach, an error model for
primitive gates is assumed and a corresponding method of quantum
error correction or avoidance is chosen. Fault tolerance conditions
are derived through a worst-case analysis of error propagation, or
by making assumptions that limit error propagation.

In this paper we propose a very different ``top-down" approach that
starts from the fault tolerance design objective, and develops
specifications for lower level functionality. Our methods enable selection of optimal quantum error correction techniques by avoiding overly pessimistic worst case analysis, or unnecessary assumptions limiting error propagation. Our approach imposes a
system-level criterion based on specification of the full system
dynamics at each level of concatenation, thus allowing for the
systematic determination of optimal error correction, avoidance and
fault-tolerance solutions. As we will show, a crucial advantage of
our top-down approach is that it leads to more accurate
determinations of error thresholds than can be obtained using the
standard, bottom-up approach. Our top-down approach is based on an
operator theoretic measure of implementation inaccuracy, which
compares the full system dynamics of the {\em actually implemented}
({\it i.e.\/}, noisy) quantum computation directly with the {\em
ideally defined} ({\it i.e.\/}, noiseless), quantum computation.
This comparison is expressed by the Quantum Computer Condition
(QCC), an explicit mathematical constraint which requires that the
difference between the actually implemented and ideally defined
computations be bounded by a parameter with a prescribed value
\cite{GHW}. The QCC provides a universal framework for quantum
computational fault tolerance that, concomitant to providing a means
to more accurately determine error threshold values than are
obtained with standard quantum fault tolerance, allows the
determination of optimal error correction codes (or error avoidance
methods), obviating the need to make {\it ad hoc} choices. This is
related to the fact that, as we show, the QCC incorporates, as
special cases, all known methods of quantum error correction and
avoidance. In the process of demonstrating the latter we introduce a
new family of error correction codes and avoidance methods.

\section{A Universal Framework for Quantum Error Correction and Avoidance}

In this section we show that the known methods of error correction
and avoidance are all special cases of a universal framework
provided by the QCC. To do this we first briefly describe and review
the known methods of error correction and avoidance, and then
introduce a new, operator theoretic formulation of entanglement
assisted error correction which contains the previously known
entanglement assisted error correction, operator quantum error
correction and standard quantum error correction protocols as
special cases. The entanglement assisted operator error correction
presented here supports a new family of error correction codes and
error avoidance methods. We then show that the entanglement assisted
\emph{operator} error correction introduced here is a special case
of the QCC. After summarizing the known methods of quantum error
correction and avoidance we review the definition of the QCC
\cite{GHW} and show that all of the methods of error correction and
avoidance described here are subsumed under the universal, operator
theoretic framework provided by the QCC.

{\it Quantum error correction}. Quantum error correction (QEC) is a
method of detecting and correcting errors that affect quantum
information \cite{QEC}. The quantum information to be protected is
stored in a subspace of a quantum system, known as the code
subspace. Errors that can be corrected by a quantum error correction
code take the state out of the code subspace. Measurement of the
error syndrome identifies the required recovery operation without
affecting the encoded quantum information. Thus, the proper
correction scheme can be applied without causing disturbance. We can
write the general error correction condition of a QEC as
\begin{equation}\label{qec}
V_{dec}R\varepsilon V_{enc} \rho  = \rho
\end{equation}
for all density matrices $\rho$, where $\rho$ contains the quantum
information to be protected.  For simplicity, we take $\rho$ to be
defined on the space of $k$ logical qubits. The map $V_{enc}$
adjoins an $s$-qubit ancilla state to $\rho$ and performs the
encoding into the code subspace of the full Hilbert space of $n=k+s$
qubits. The dynamical map $\varepsilon$ describes the errors that
affect the encoded state. $R$ is the syndrome measurement and
recovery operation.  The map $V_{dec}$ is the decoding operation
corresponding to $V_{enc}$, which yields the final state consisting
of $k$ logical qubits.  Error correction is thus successful if this
final state is the same as the initial state $\rho$.

{\it Operator quantum error correction}. Operator quantum error
correction (OQEC) \cite{Kribs} is a superoperator formalism that
generalizes QEC, incorporating  error avoidance techniques,
including decoherence free subspaces and noiseless subsystems, into
a single framework. Following \cite{Kribs}, OQEC partitions the
system's Hilbert space as $H = A \otimes B \oplus K$. Information is
stored on subsystem $A$ in the encoded state $\rho_A = V_{enc}\rho$,
where $\rho$ is the state of $k$ logical qubits and $\rho_A$ is the
state of $n=k+s$ qubits as for QEC. The full OQEC state defined on
the complete Hilbert space, $H$, is obtained {\emph {via}}
$W_{\rho_B}(\rho_A)\equiv\rho_A \otimes \rho_B \oplus 0_{K}$, where
$\rho_B$ is an arbitrary density matrix defined on space $B$. Due to
error avoidance techniques, certain errors may only affect $\rho_B$.
Errors that are not thus protected will affect $\rho_A$, and are
described by error map $\varepsilon$. These errors are deemed
correctable if there exists a dynamical map $R$ that reverses the
action of $\varepsilon$ on $\rho_A$, up to a transformation on space
$B$. We can write the general error correction condition of OQEC as
\begin{equation}\label{oqec}
V_{dec}Tr_B F_{AB}R\varepsilon W_{\rho _B }V_{enc}\rho  = \rho
\end{equation}
for all $\rho$, where $F_{AB}$ is a projection of $H$ onto $A
\otimes B$ and $Tr_B$ is the partial trace over $B$. Standard QEC
codes are thus a special case of OQEC where the dimension of $B$ is
$1$.

{\it Entanglement assisted quantum error correction}. Entanglement
assisted quantum error correction (EAQEC) is another generalization
of QEC \cite{Brun06}. Error correction in general pertains to the
protection of quantum information as it traverses a quantum channel.
We will refer to the transmitter and receiver, respectively, as
Alice and Bob. Alice would like to encode $k$ logical qubits into
$n$ physical qubits which will be sent to Bob {\it via} a noisy
channel, $\varepsilon$. In EAQEC Alice and Bob are presumed to share
$c$ maximally entangled pairs of qubits. Alice encodes her $k$
qubits into $n = k + c + s$ qubits {\it via} the operation ${\cal
V}_{enc}$, where $s$ is the number of required ancilla qubits, and
we have replaced the symbol $V_{enc}$ used in QEC and OQEC with the
symbol ${\cal V}_{enc}$ to indicate that the size of the target
space has been increased from $k+s$ to $k+c+s$. Bob measures an
error syndrome using his portion of the initially shared ebits in
addition to the $n$ qubits he received from Alice.  He then carries
out any transformations required to recover from errors. We denote
the syndrome measurement followed by the recovery with the symbol
${\cal R}$, replacing the symbol $R$ used for QEC and OQEC, to
reflect the fact that the operation now involves the additional
shared ebits. Finally, he performs the decoding operation ${\cal
V}_{dec}$ to obtain a final state of $k$ logical qubits. Error
correction is successful if this is the same as Alice's initial
state of $k$ qubits.  Following \cite{Brun06}, we say that an
$[[n,k;c]]$ EAQEC code consists of an encoding operation ${\cal
V}_{enc}$ and a decoding operation ${\cal V}_{dec}$. The error
correction condition can then be written as
\begin{equation}\label{eaqec}
{\cal V}_{dec} {\cal R}\varepsilon {\cal V}_{enc} \rho  = \rho~.
\end{equation}
This differs from the QEC condition (eq.(\ref{qec})) in that the
encoding and decoding operations ${\cal V}_{enc}$ and ${\cal
V}_{dec}$, as well as ${\cal R}$, act on the initially shared ebits
in addition to the ancilla. We obtain QEC as a special case of EAQEC
by eliminating the ebits, that is, by setting $c=0$.


{\it Entanglement assisted operator quantum error correction}. We
now generalize the EAQEC of \cite{Brun06} by introducing and
defining an operator theoretic generalization of it that we shall
refer to as entanglement assisted operator quantum error correction
(EAOQEC), that includes EAQEC, as well as OQEC, and hence QEC, as
special cases. We then show that EAOQEC, and thus all known methods
of quantum error correction and avoidance, are special cases of the
QCC. The EAOQEC coding and decoding scheme is summarized in the
following steps: (1) ${\cal V}_{enc}$, encodes Alice's state $\rho$
to $\rho_A$
using the $c$ shared ebits and $s$ ancilla qubits. (2) Alice encodes
$\rho_A$ as a generalized noiseless subsystem into a larger system
${\cal W}_{\rho_B}(\rho_A)= \rho_A \otimes \rho_B \oplus 0_K$,
replacing the symbol $W_{\rho_B}$ used for OQEC, reflecting the
presence of shared ebits. (3) Alice sends the qubits through a noisy
channel, $\varepsilon$. (4) Bob performs syndrome measurement and
recovery operation ${\cal R}$. (5) ${\cal F}_{AB}$ projects the
entire Hilbert space onto $A \otimes B$, replacing the symbol
$F_{AB}$ used for OQEC to reflect the fact that the projection
applies to an enlarged space involving the additional shared ebits.
(6) The noisy part of the encoded system is traced out as
$Tr_B(\rho_A \otimes \rho_B)= \rho_A$. (7) ${\cal V}_{dec}$, the
final decoding, recovers the initial state. The EAOQEC formulation
presented here allows for new error correction and avoidance
protocols beyond those already incorporated in QEC, OQEC, and EAQEC.
The error correction condition of EAOQEC is given by
\begin{equation}\label{eaoqec}
{\cal V}_{dec} Tr_B {\cal F}_{AB} {\cal R}\varepsilon {\cal W}_{\rho
_B } {\cal V}_{enc} \rho  = \rho ~.
\end{equation}
We see that EAQEC codes arise as a special case of EAOQEC, {\it
i.e.}, eq.(\ref{eaoqec}) reduces to eq.(\ref{eaqec}), when $\dim
B=1$. We see that OQEC emerges as a special case of EAOQEC, {\it
i.e.}, eq.(\ref{eaoqec}) reduces to eq.(\ref{oqec}), when $c=0$
(note that $\lim_{c\rightarrow 0}{\cal V}_{{enc\atop
dec}}=V_{{enc\atop dec}}$, $\lim_{c\rightarrow 0}{\cal
F}_{AB}=F_{AB}$, $\lim_{c\rightarrow 0}{\cal R}=R$ and
$\lim_{c\rightarrow 0}{\cal W}_{\rho _B }=W_{\rho _B }$). Finally,
we see that we obtain QEC as a special case of EAOQEC, {\it i.e.},
reduce eq.(\ref{eaoqec}) to eq.(\ref{qec}), when $\dim B=1$ and
$c=0$ \cite{f1}. In the next sub-section of the paper, we show that
EAOQEC is itself a special case of the QCC.

\subsection{The Quantum Computer Condition}

The QCC compares the dynamics of the initial quantum state $\rho$
under the unitary evolution of an ideal (noiseless) quantum
computer, $U$, to the evolution of the state under the actual
implementation with a real (noisy) quantum computer $P$, by taking
their difference under the Schatten-1 norm:
\begin{equation}\label{encoded_QCC}
\|\mathcal{M}_{\{\mathrm{c}\rightarrow\mathrm{l}\}}(P\cdot(\mathcal{M}_{\{\mathrm{l}\rightarrow\mathrm{c}\}}(\rho)))
- U\rho U^\dag\|_1 \leq \alpha.
\end{equation}
The left hand side, which we call the \emph{implementation
inaccuracy}, is the norm of the difference between the actual
(implemented) and desired (ideal) final quantum states. It is
bounded by $\alpha$, a prescribed accuracy required of the quantum
computation. Because the ideal (unitary) operator $U$ acts on {\em
logical} qubits and the actual (positive dynamical map \cite{f2})
implementation $P$ acts on the {\em computational} or physical
qubits, many of which may be needed to encode one logical qubit, we
introduce the pair of linking maps,
$\mathcal{M}_{\{\mathrm{c}\rightarrow\mathrm{l}\}}$ and
$\mathcal{M}_{\{\mathrm{l}\rightarrow\mathrm{c}\}}$ to connect the
logical and computational Hilbert spaces \cite{GHW}.  Note that
these maps do {\em not} represent the physical processes which
encode and decode qubits.  All such physical processes are, in
principle, subject to noise and are therefore properly part of the
dynamics represented by $P$ \cite{f3}.  It is convenient to define
$\tilde{P} \equiv \mathcal{M}_{\{\mathrm{c}\rightarrow\mathrm{l}\}}
\cdot P \cdot \mathcal{M}_{\{\mathrm{l}\rightarrow\mathrm{c}\}}$, in
terms of which the QCC becomes
\begin{equation}\label{M_subsumed_QCC}
\|\tilde{P} (\rho) - U\rho U^\dag\|_1 \leq \alpha.
\end{equation}
We take note of the fact that, in considering the properties of the
implementation inaccuracy (\emph{i.e.}, the left-hand-side of either
eq.\ref{encoded_QCC} or \ref{M_subsumed_QCC}), its origin in an
inequality that must hold for all states, $\rho$, implicitly
ascribes the operation of taking the supremum over the complete set
of density matrices.

The QCC concisely incorporates a complete specification of the full
dissipative, decohering dynamics of the actual, practical device
used as the quantum computer, a specification of the ideally-defined
quantum computation intended to be performed by the computer, and a
quantitative criterion for the accuracy with which the computation
must be executed given the inevitability of residual errors
surviving even after error correction has been applied.  Thus, the
QCC provides a rigorous universal, high level framework allowing for
the top-down analysis of fault tolerant quantum computation.

\begin{figure}
\includegraphics[height=4cm]{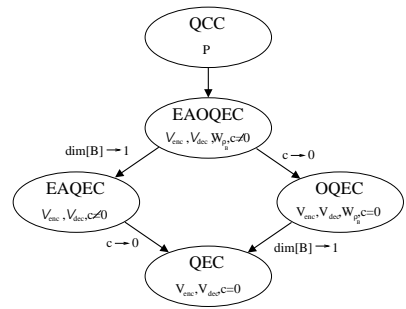}
\caption{The relationship between the categories of error correction and avoidance 
is illustrated in this figure, as discussed in detail in the text.  The arrows 
go from general cases to special cases, illustrating that all forms of 
error correction and error avoidance are special cases of the QCC.    }
\label{fig:qechierarchy}
\end{figure}

As a universal framework for quantum computation we now show that
the QCC incorporates EAOQEC, and thus OQEC, EAQEC, and QEC as well,
as special cases. (In the following section, we demonstrate that the
QCC goes beyond quantum error correction and provides a
superoperator approach to quantum fault-tolerance.) The error
correction condition of EAOQEC is given by:
\begin{equation}
{\cal V}_{dec} Tr_B {\cal F}_{AB} {\cal R}\varepsilon {\cal
W}_{\rho_B } {\cal V}_{enc} \rho  = \rho
\end{equation}
This can be rewritten as:
\begin{equation}
||{\cal V}_{dec} Tr_B {\cal F}_{AB} {\cal R}\varepsilon {\cal
W}_{\rho_B } {\cal V}_{enc} \rho - \rho ||_1 = 0,
\end{equation}
which is a special case of the QCC with $\alpha=0$, $U=I$ (we note
that quantum error correction in general protects quantum
information that traverses a quantum communications channel, and
that such a channel can be regarded as a quantum computer that
implements the identity operation), and $\tilde{P} = {\cal V}_{dec}
Tr_B {\cal F}_{AB} {\cal R} \varepsilon {\cal W}_{\rho _B } {\cal
V}_{enc}$. (Note that the noiseless linking maps
$\mathcal{M}_{\{\mathrm{c}\rightarrow\mathrm{l}\}}$ and
$\mathcal{M}_{\{\mathrm{l}\rightarrow\mathrm{c}\}}$ are subsumed in
the definitions of ${\cal V}_{enc}$ and ${\cal V}_{dec}$.) Thus
EAOQEC is a special case of the QCC, and, given the discussion
immediately following eq.(\ref{eaoqec}), we see that all known
protocols for quantum computational error correction and avoidance
therefore arise as special cases of the QCC as well. The
relationship among the different categories of error correction is
depicted in Figure \ref{fig:qechierarchy}.

\section{Operator Quantum Fault Tolerance (OQFT): A Universal Framework}

In the previous section we showed that the QCC provides a universal
framework that encompasses all of the known approaches to quantum
error correction and avoidance.  In this section we apply the QCC to
the further problem of quantum fault tolerance.  We will show that
the use of the QCC leads naturally to a top-down approach to the
problem of quantum computational fault tolerance. In this top-down
approach, the success criterion for fault tolerance is expressed in
terms of operator norms that characterize the difference between the
quantum state produced by the actual quantum computer and the
desired quantum state which would be produced by a perfect,
noiseless quantum computer. We refer to this top-down approach as
Operator Quantum Fault Tolerance (OQFT) by virtue of the fact that
the approach is based on operator theory. OQFT thus generalizes
standard quantum fault tolerance theory in a manner analogous to the
generalization of QEC by OQEC. Our approach is to be contrasted with
the standard, bottom-up approach in which the success criterion is
expressed in terms of derived error probabilities, without reference
to a quantification of the effects of the errors on the quantum
state itself. An advantage of the top-down approach is that it
allows for more accurate computation of error-thresholds.

The quantum error correction and avoidance techniques discussed in
the previous section are designed to protect quantum information as
it traverses a quantum {\it channel}. Thus, in the statement of the
QCC, the desired unitary operation is simply the identity $U=I$.
Quantum fault tolerance applies to the more general case of quantum
{\it computation}, where $U\neq I$ in general. Quantum fault
tolerance also allows for the possibility that errors may arise
during the execution of the quantum error correction/avoidance
algorithm itself. The presence of a non-zero $\alpha$ in the right
hand side of the QCC corresponds to the realization that, despite
our encoding efforts, there is always some residual error in the
final state of the quantum computation. Thus, the QCC provides a
proper framework for the study of quantum fault tolerance, the
procedure of producing the (nearly) correct final quantum state.

In order to deduce the conditions required for fault-tolerance, we
proceed by examining the consequences of concatenating a quantum
error-correction code or an error-avoidance method, labeling the
successive levels of concatenation by the index $(i)$. At each level
of concatenation, the implementation inaccuracy is given by
\begin{equation}
\|\tilde{P}^{\left( i \right)} \rho - U \rho U^\dag\|_1  ~.
\end{equation}
We may now write down the condition that comprises OQFT. Fault
tolerance is deemed successful at a given level of concatenation if
the implementation inaccuracy at concatenation level $i+1$ is less
than the implementation inaccuracy at level $i$, that is, if
\begin{equation}\label{sup_bound}
\frac {\opr{sup}_\rho \|\tilde{P}^{\left( i+1 \right)}\rho - U \rho
U^\dag\|_1}
 {\opr{sup}_\rho \|\tilde{P}^{\left( i \right)}\rho - U \rho U^\dag\|_1} < 1~,
\end{equation}
where the supremum operation arises in accordance with the
discussion immediately following eq.(\ref{M_subsumed_QCC}). The OQFT
condition, the inequality appearing in (\ref{sup_bound}), is a
completely general, operator-theoretic condition that must be
satisfied in order to achieve fault tolerant operation of a quantum
computer \cite{noncircuit}. This is a top-down condition, obtained within the
framework provided by the QCC, that allows for more accurate
determination of error thresholds than is possible in the standard,
bottom-up approach, as we will see in the following example. The
example also illustrates that the standard bottom-up approach to
fault tolerance arises as a special case of the OQFT condition given
in (\ref{sup_bound}).

\subsection{Application of the QCC to Fault Tolerance}

We illustrate the application of eq.(\ref{sup_bound}) to fault
tolerance with the following example. We first introduce an
operator, $\Upsilon$, presumed to act on the computational Hilbert
space, that faithfully implements the effect of the unitary operator
$U$ (which acts on the smaller, logical Hilbert space):
\begin{equation}\label{EQ:perfectcircuit}
\mathcal{M}_{\{\mathrm{c}\rightarrow\mathrm{l}\}} \left[ \Upsilon
              \left( \mathcal{M}_{\{\mathrm{l}\rightarrow\mathrm{c}\}}\rho \right)
             \Upsilon^\dag \right] = U \rho U^\dag.
\end{equation}
We now introduce a dynamical model for the evolution of the state of
the quantum computer, at a given level of concatenation $i$,
including computational errors, given by:
\begin{eqnarray}
\label{EQ:errormodel} {P^{\left( i \right)}} \left(
\mathcal{M}_{\{\mathrm{l}\rightarrow\mathrm{c}\}}\rho \right) &=&
   \left( 1 - \epsilon^{(i)} \right)
             \Upsilon \left( \mathcal{M}_{\{\mathrm{l}\rightarrow\mathrm{c}\}}\rho \right) \Upsilon^\dag \nonumber\\
   && \phantom{aa} +
   \epsilon^{(i)} Q^{\left( i \right)} \left( \mathcal{M}_{\{\mathrm{l}\rightarrow\mathrm{c}\}}\rho \right) ~,
\end{eqnarray}
where $\epsilon^{(i)}$ depends on the probability for the occurrence of errors
and $Q^{\left( i \right)}$ is a superoperator that represents the
effects of the errors in the computational Hilbert space. Note that
the superoperator, $P^{\left( i \right)}$, is a linear combination
of the operator, $\Upsilon$, and the superoperator, $Q^{\left( i
\right)}$. The above model used in the present example is
sufficiently general to include, as special cases, local stochastic
noise and locally correlated stochastic noise, the models that
define standard quantum fault-tolerance \cite{std_qft}. The
implementation inaccuracy is then
\begin{equation}
\begin{aligned}
\|\tilde{P}^{\left( i \right)}\rho & - U \rho U^\dag\|_1 = \\
 &  {\Big \|} \left( 1 - \epsilon^{(i)} \right)
           \mathcal{M}_{\{\mathrm{c}\rightarrow\mathrm{l}\}} \left[
              \Upsilon \left( \mathcal{M}_{\{\mathrm{l}\rightarrow\mathrm{c}\}}\rho \right)
              \Upsilon^\dag \right] \\
 & \;\;\;\;\;\;\;\; + \epsilon^{(i)}
              \tilde{Q}^{\left( i \right)} \rho - U \rho U^\dag {\Big \|}_1 ~,
\end{aligned}
\end{equation}
where we have absorbed the linking maps into $\tilde{Q}^{\left( i
\right)} \equiv\mathcal{M}_{\{\mathrm{c}\rightarrow\mathrm{l}\}}
Q^{\left( i \right)}
\mathcal{M}_{\{\mathrm{l}\rightarrow\mathrm{c}\}}$. With
(\ref{EQ:perfectcircuit}) the implementation inaccuracy then reduces
to the simple form
\begin{equation}
\|\tilde{P}^{\left( i \right)}\rho - U \rho U^\dag\|_1 = \epsilon^{(i)}
   \|  \tilde{Q}^{\left( i \right)} \rho - U \rho U^\dag  \|_1 ~.
\end{equation}
The OQFT success criterion given by eq.(\ref{sup_bound}) then
becomes
\begin{equation}
\label{EQ:epsnormratio} \frac {\epsilon^{\left( i+1
\right)}}{\epsilon^{\left( i \right)}}
   \cdot \frac {\opr{sup}_\rho \|  \tilde{Q}^{\left( i+1 \right)} \rho - U \rho U^\dag  \|_1}
               {\opr{sup}_\rho \|  \tilde{Q}^{\left( i \right)} \rho - U \rho U^\dag  \|_1} < 1~.
\end{equation}
Note that this differs in an important way from the criterion used
in standard treatments of fault tolerance, which is based {\it
solely on a comparison of error probabilities}:
\begin{equation}
\label{EQ:standardqft} \frac {\epsilon^{\left( i+1
\right)}}{\epsilon^{\left( i \right)}}
    < 1~.
\end{equation}
Equation (\ref{EQ:epsnormratio}) {\it takes explicitly into account
the fact that the error dynamics associated with residual errors in
general differs at each level of concatenation}, namely, that
$Q^{\left( i+1 \right)} \neq Q^{\left( i \right)}$. If the
dynamical effects of the errors tend to become smaller at higher
levels of concatenation, quantum fault tolerance can be achieved at
error thresholds larger than those predicted by standard fault
tolerance theory. This improvement in error threshold values arises
due to the correction provided by the ratio of dyamical factors
derived from the implementation inaccuracy.

\subsection{Example: OQFT condition with five-qubit code}

As a concrete example of Eqs.~10 and 15 we analyze a case where we attempt
to store one qubit of information, and thus have $U = I$, using an error
correction code. We assume a general environment with two independent noise
generators: one causing bit-flips with probability $p_x$, and the
other phase-flips with probability $p_z$. For simplicity we assume
perfect readout and recovery operations. We note that the
case of unbalanced noise generators ({\em {i.e.}}, biased noise) is physically
important, and achieving fault tolerance in such an environment has been
the subject of recent analysis \cite{pajp}.
The final state of the one qubit of information after encoding,
occurrence of errors, and any attempts at recovery, and decoding if
necessary, will be:
\begin{equation}
\label{ex} {\tilde P}^{(i)}\rho =
{\Big (}1-\epsilon^{(i)}{\Big )}\rho+
\epsilon^{(i)}{\Big (}\eta_x^{(i)}\rho_x+\eta_y^{(i)}\rho_y+\eta_z^{(i)}\rho_z{\Big )},
\end{equation}
where $\rho_k = \sigma_k\rho\sigma_k^\dagger$ for $k = x,y,z$,
the $\sigma_k$ are the Pauli operators,
$\eta_k^{(i)} \equiv \epsilon_k^{(i)}/\epsilon^{(i)}$,
$\epsilon^{(i)} \equiv \sum_{k}\epsilon_k^{(i)}$, and the
$\epsilon_k^{(i)}$ are functions of the concatenation level, $i$,
the noise generators represented by $p_x$ and $p_z$, the error
correcting code, and the recovery operation. We emphasize
the dependence of $\epsilon_k^{(i)}$ on the concatenation level,
noting that if the concatenation procedure is working properly,
the quantity $\epsilon^{(i)}$ should decrease with
increasing $i$.

We now calculate the supremum of the implementation inaccuracy at the $i$th
concatenation level using Eq. \ref{ex}:
\begin{equation}
\label{ii} \opr{sup}_\rho \| {\tilde P}^{(i)}\rho - \rho \|_1 =
\epsilon^{(i)} \opr{sup}_\rho \| \eta_x^{(i)}\rho_x+\eta_y^{(i)}\rho_y+\eta_z^{(i)}\rho_z-\rho \|_1.
\end{equation}
The OQFT success
criterion, eq.(\ref{sup_bound}), for applying concatenation up to level $i+1$, is now
\begin{equation}
\label{iisuccess}
\frac{\epsilon^{(i+1)}}{\epsilon^{(i)}}\cdot\frac{\opr{sup}_\rho \| \eta_x^{(i+1)}\rho_x+\eta_y^{(i+1)}\rho_y+\eta_z^{(i+1)}\rho_z-\rho \|_1}
{\opr{sup}_\rho \| \eta_x^{(i)}\rho_x+\eta_y^{(i)}\rho_y+\eta_z^{(i)}\rho_z-\rho \|_1}<1.
\end{equation}
It is important to note that the quantities $\eta_k$ appearing in eq.(\ref{iisuccess}) do not in general factor out of the norm. The first factor on the left-hand-side of eq.~(\ref{iisuccess}) is analogous to that of standard quantum fault tolerance
({\em {cf}} ~eq.(\ref{EQ:standardqft})), while the second factor on the left-hand-side of eq.~(\ref{iisuccess}) is the {\em correction factor} that arises upon imposing the condition of operator quantum fault tolerance. For simplicity, we will refer to the ratio comprising the entire left-hand-side of eq.~(\ref{iisuccess}) as the OQFT $P$-ratio ({\em cf} ~eq.(\ref{sup_bound})). We will refer to the ratio of norms on the left-hand-side of eq.~(\ref{iisuccess}) that appears to the right of the dot as the OQFT $Q$-ratio ({\em cf} ~eq.(\ref{EQ:epsnormratio})). We will refer to the ratio appearing in eq. (\ref {EQ:standardqft}) as the (standard) quantum fault tolerance (QFT) ratio.

\begin{figure}
\includegraphics[width=4.274cm]{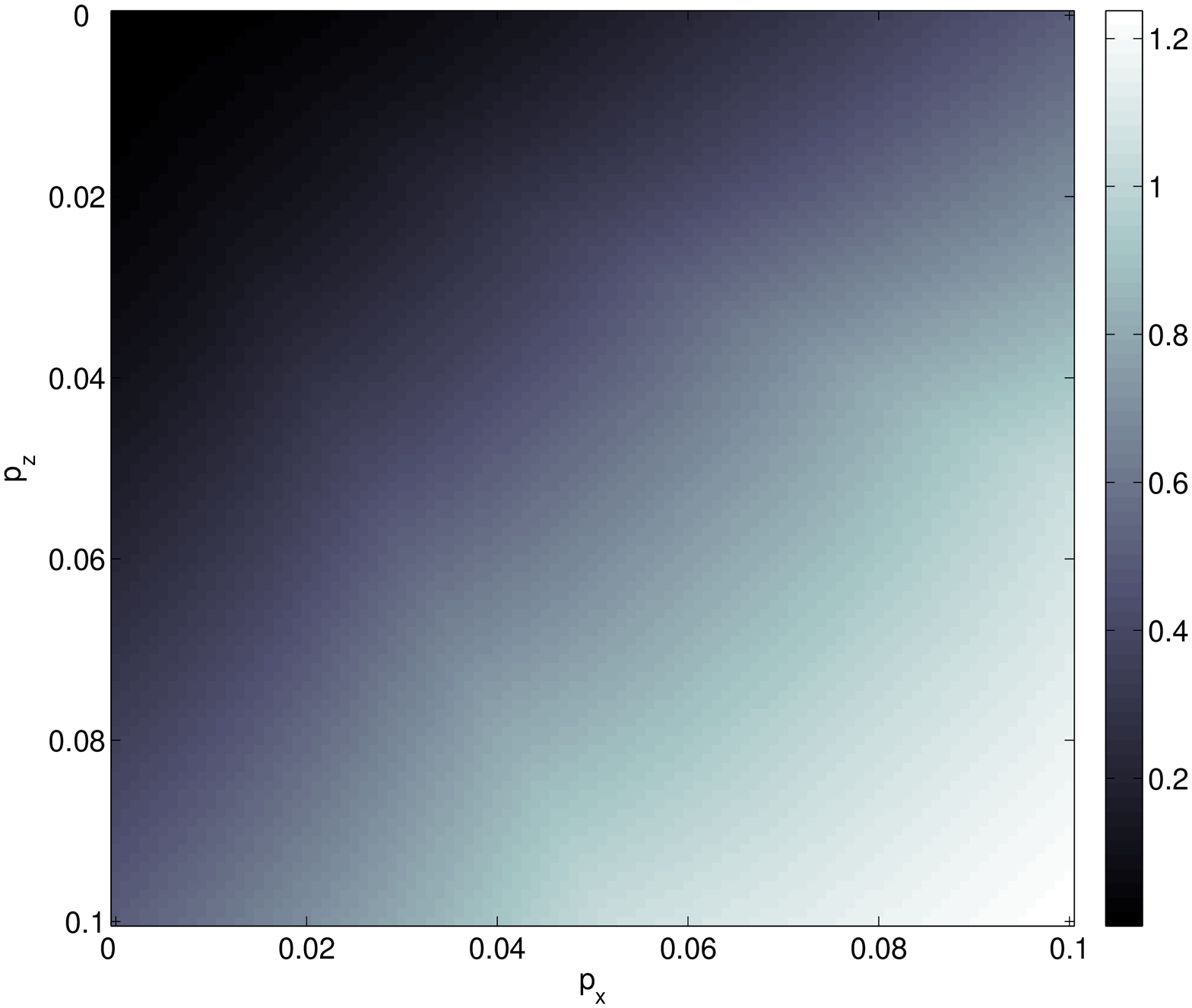}
\includegraphics[width=4.274cm]{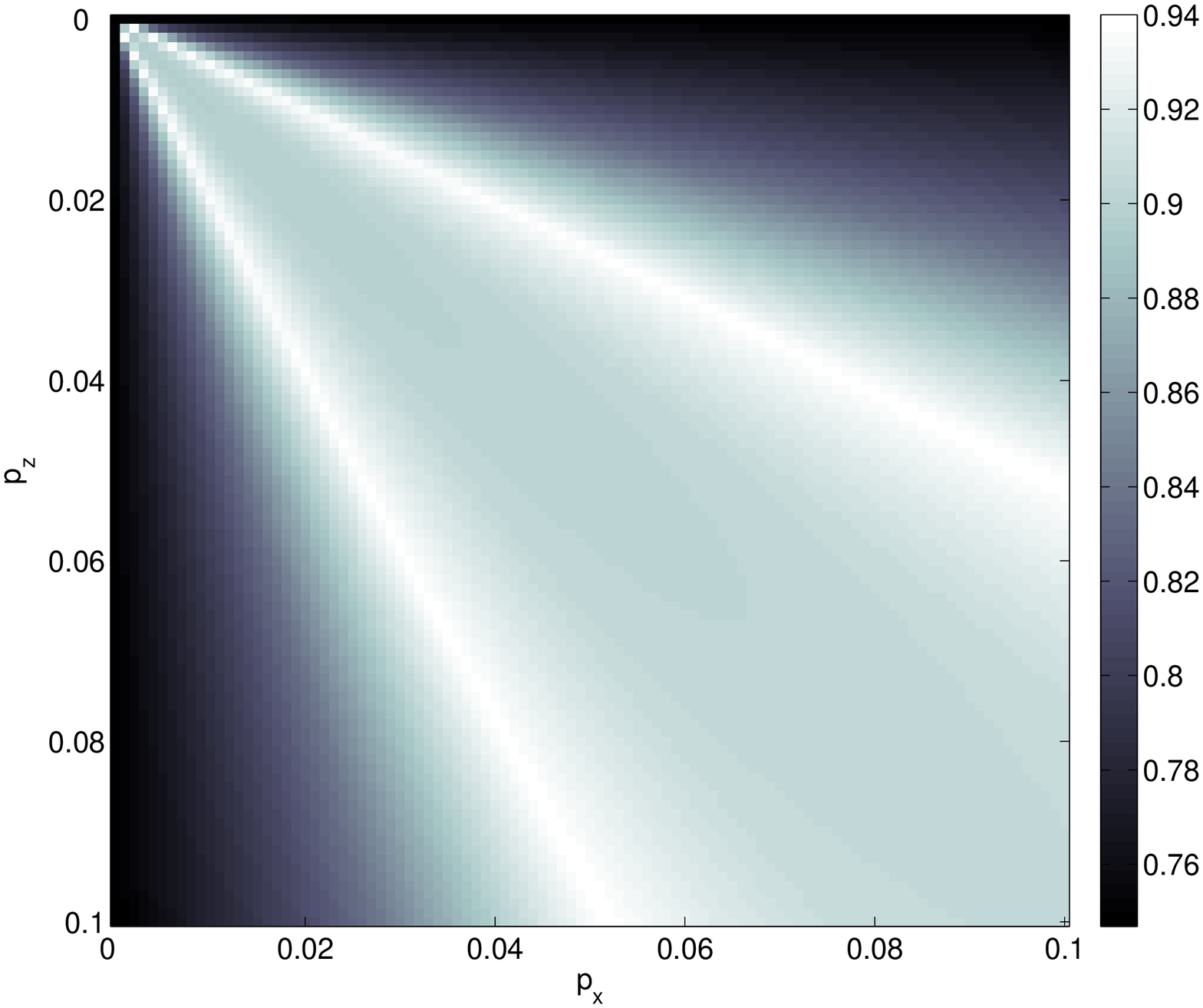}
\caption{$P$-ratio (left) and $Q$-ratio (right), when comparing the second
level of concatenation to the first level of concatenation, for the [5,1,3] quantum error correction code protecting one qubit of quantum information. The $x$ and $y$ axes are the values of $p_x$ and $p_z$. If the $P$-ratio is greater than 1 the concatenation is unsuccessful. The $Q$-ratio, correction term to the standard
approach to quantum fault-tolerance, is shown here to be always less
than 1, implying that standard QFT threshold estimates are too tight. }
\label{q5}
\end{figure}

Proceeding with the analysis, we now assume an environment which causes bit flips to occur with probability $p_x$, and phase flips with probability $p_z$. We will employ the [5,1,3] code to protect one qubit of quantum information and compare one, and
two, concatenations of the code, making use of the OQFT condition (eq.(\ref{sup_bound})
or eq.(\ref{iisuccess})) as the criterion for fault tolerance. The numerical results of this analysis for a region in \{$p_x$, $p_z$\!\} space are shown in Fig.~\ref{q5}. Consider the graph on the left in Fig.~\ref{q5}. Those data points in the graph for which the $P$-ratio vaues are less than $1$ correspond to scenarios for which the concatenation is successful. In contrast, those points in the graph for which the $P$-ratio values are greater than $1$ correspond to scenarios for which the concatenation to this level is unsuccessful, {\emph {i.e.}}, concatenation will increase errors. The $Q$-ratio, displayed in the graph on  the right in Fig.~\ref{q5}, which provides the correction factor to the standard fault-tolerance analysis, is always less than $1$ in the current example. Thus, we see for this example that the full OQFT analysis presented in this paper yields thresholds that are fine-tuned to the physical environment, and are thus more accurate ({\em {i.e.}}, less stringent) than thresholds arising from standard QFT analysis.

\begin{figure}
\includegraphics[width=8cm]{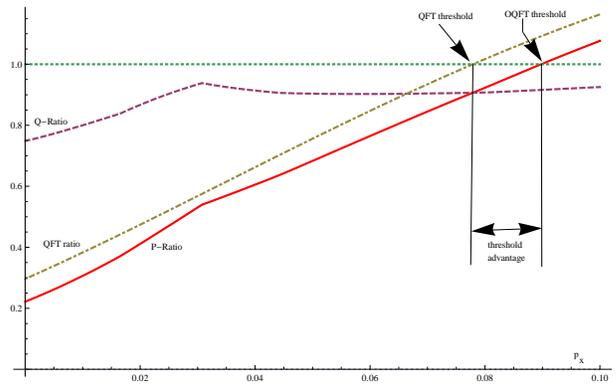}
\caption{$P$ (solid line), $Q$ (dashed line), and QFT (chain) ratios when comparing the second level of concatenation to the first level of concatenation, for the [5,1,3] quantum error correction code protecting one qubit of quantum information. We have set $p_z$ to .06. The $Q$-ratio is always less than $1$ (dotted line) and therefore the $P$-ratio will be smaller than the QFT-ratio. For the example shown in the plot, the standard quantum fault tolerance condition would indicate that concatenation fails when $p_x \agt .0785$ because at that point the QFT ratio is greater than one. The corresponding OQFT threshold value is about 15\% higher, showing that concatenation actually fails only when $p_x \agt .0905$. } \label{qratios}
\end{figure}

Pursuing this further, Figure \ref{qratios} shows the various ratios of interest for a case where we have set $p_z$ to .06, which would correspond to a situation of practical experimental interest. In this figure we highlight the
effect of the $Q$-ratio on the threshold value. Specifically, using the
standard QFT ratio ({\em cf} ~eq.(\ref{EQ:standardqft})), concatenation would have been deemed unsuccessful for any $p_x \agt .0785$ since at that point the
QFT-ratio becomes greater than $1$. The correction factor of OQFT
relaxes this value, and we see that, in fact, concatenation to the second level is successful for $p_x$ values as large as $p_x \alt .0905$ (at which point the $P$-ratio becomes greater than one). Thus, by enabling fine tuning to the physical environment, the OQFT condition yields a {\em {threshold advantage}} of about $15\%$ compared to the result obtained from standard QFT.

\section{Conclusions}

In this paper we introduced operator quantum fault tolerance (OQFT),
a universal, operator-theoretic framework for quantum fault
tolerance that includes previously known, standard quantum fault
tolerance as a special case. OQFT comprises a top-down approach that
employs a system-level criterion based on specification of the full
system dynamics. OQFT allows us to calculate more accurate values of
error correction thresholds than is possible using previous
approaches to fault tolerance. This was demonstrated both formally and with an explicit numerical example. OQFT is based on the quantum computer
condition (QCC) which was demonstrated to provide a universal
organizing structure for all known methods of error correction and
error avoidance. We did this by introducing a new, operator
theoretic form of entanglement assisted error correction, which
includes all the previously known methods of error correction and
avoidance, and is itself a special case of the QCC. 

\section{Acknowledgements}
GG, MH, and YSW acknowledge support from the MITRE Technology
Program under MTP grant \#07MSR205. VA and ARC acknowledge partial
support from the National Science Foundation under NSF grant
\#1096066, and from the Air Force Office of Scientific Research
under contract \#00852833. We thank M. Wilde, and the authors of \cite{newbrun},
for kindly bringing \cite{newbrun} to our attention.

\section{Appendix}
After the present paper was completed, we were apprised of an
independent formulation of EAOQEC presented by Hsieh, {\it et al.}
in \cite{newbrun}. The connection between the two formulations can
be made transparent by considering an equation that depicts both:
\begin{equation}
\label{eeaoqec} \mathcal{V}_{dec,kin}{\Big
\{}{\mathcal{V}_{dec,dyn}\cdot\mathcal{R} \atop
\tilde{\mathcal{R}}\cdot\tilde{\mathcal{V}}_{dec,dyn}}{\Big
\}}\varepsilon\mathcal{V}_{enc,dyn}\mathcal{V}_{enc,kin}\cdot\rho =
\rho.
\end{equation}
Here the upper line within curly brackets corresponds to the EAOQEC
of the present paper, and the lower line corresponds to that of
\cite{newbrun}. In this equation, we have factored the encoding and
decoding operations into separate kinematic and dynamic parts. The
kinematic factors append or remove any necessary ancilla qubits and
ebits ({\it i.e.}, increase or decrease, as required, the Hilbert
space dimension), and the dynamic factors represent unitary
evolution of the qubits ({\it i.e.}, implement the physical encoding
and decoding operations) \cite{commute}.

The principal difference between the EAOQEC formulation presented in
the present paper and that presented in \cite{newbrun} is seen in
the placement of the unitary part of the decoding operation relative
to the measurement and recovery operation. In the present paper the
unitary dynamical part of the decoding operation,
$\mathcal{V}_{dec,dyn}$, is performed {\it after} a (general)
measurement and recovery operation, $\mathcal{R}$, whereas in
\cite{newbrun} the unitary decoding operation,
$\tilde{\mathcal{V}}_{dec,dyn}$ (represented by the symbol $U$ in
\cite{newbrun}), is performed {\it before} a (specifically defined)
measurement and recovery operation, $\tilde{\mathcal{R}}$
(represented by the symbol $\mathcal{D}_0$ in \cite{newbrun}). Thus,
the relative order of the unitary decoding operation and the
measurement/recovery operation differs between the two formulations
of EAOQEC. We note that the unitary decoding operation and the
measurement/recovery operation do not in general commute. We further
note that the formulation of EAOQEC given in \cite{newbrun} is
defined within the stabilizer formalism, whereas the formulation of
EAOQEC given in the present paper allows for non-stabilizer codes,
such as non-additive codes, in addition to stabilizer codes.

Finally, we recapitulate that we have shown that all forms of error
correction and error avoidance are in fact special cases of the more
fundamental QCC (with $\alpha = 0$ and $U=I$ in either
eq.(\ref{encoded_QCC}) or (\ref{M_subsumed_QCC})), and that the QCC
(with $\alpha > 0$ and $U$ arbitrary in either
eq.(\ref{encoded_QCC}) or (\ref{M_subsumed_QCC})) provides a
universal operator theoretic framework for quantum fault tolerance,
as demonstrated in the present paper.



\begin{thebibliography}{99}
\bibitem{f1}
In addition to EAQEC, Ref. \cite{Brun06} introduces a protocol based
on EAQEC called catalytic operator quantum error correction. It is
straightforward to show that catalytic operator quantum errror
correction is also a special case of the EAOQEC protocol that we
introduce in the current paper, and hence is a special case as well
of the QCC.
\bibitem{f2}
The QCC holds for the general case of positive dynamical maps, of
which completely-positive dynamical maps are a special case.
\bibitem{commute}
In eq.(\ref{oqec}), as described in the text, we have essentially
adopted the notation used in \cite{Kribs} (this notation with
modifications is inherited in eq.(\ref{eaoqec}) as well). In
re-casting eq.(\ref{eaoqec}) as the upper line of eq.(\ref{eeaoqec})
we have collected all kinematical operations in ${\cal
V}_{dec,~kin}$, thus conforming with the presentation given in
\cite{newbrun} (${\cal V}_{dec,~kin}$ is not represented by a symbol
in \cite{newbrun},  but is referred to as ``discarding the unwanted
systems.").
\end{thebibliography}
\end{document}